\documentclass[12pt]{article}

\usepackage{amsmath,amssymb,amsfonts} \usepackage{bm} \usepackage{braket} \usepackage{graphicx}
\usepackage{xcolor}
\usepackage{float} \usepackage[numbers]{natbib} \usepackage{hyperref} \usepackage{booktabs}
\usepackage{siunitx} \usepackage[symbol]{footmisc}

  \graphicspath{{figures/}}

\newcommand{\R}{\mathbb{R}}
\newcommand{\Sph}{\mathbb{S}}
\newcommand{\e}{\boldsymbol e}
\newcommand{\y}{\boldsymbol y}
\newcommand{\vvec}{\boldsymbol v}
\renewcommand{\v}{\boldsymbol v}
\newcommand{\w}{\boldsymbol w}
\newcommand{\cvar}{\boldsymbol c}
\newcommand{\x}{\boldsymbol x}
\newcommand{\z}{\boldsymbol z}
\newcommand{\I}{\boldsymbol I}
\newcommand{\Qrot}{\boldsymbol R}
\newcommand{\dd}{\mathrm d}
\newcommand{\Kn}{\varepsilon}
\renewcommand{\a}{\boldsymbol a}
\renewcommand{\b}{\boldsymbol b}
\renewcommand{\c}{\boldsymbol c}
\newtheorem{remark}{Remark}[section]

\makeatletter
\AtBeginDocument{%
  \let\revision@lbibitem\@lbibitem
  \def\@lbibitem[#1]#2{%
    \color{black}%
    \ifcsname revisionbib@#2\endcsname\color{red}\fi
    \ifcsname bluebib@#2\endcsname\color{blue}\fi
    \revision@lbibitem[#1]{#2}%
  }%
}

\makeatother

\begin{document}
	
	\begin{center}
		{\Large\bf Kinetic Reduction and Hydrodynamics under Time-Dependent Helical Symmetry}\par\vspace{6pt}
		Qingyun Zeng and Fan Feng\footnote[1]{Corresponding author: fanfeng@pku.edu.cn}\par
		\small School of Mechanics and Engineering Science, Peking University, Beijing 100871, China
	\end{center}
	
	\begin{abstract}
		Translational and helical symmetries provide idealized descriptions of many flows encountered in nature and engineering. While translationally symmetric flows have been extensively studied within kinetic and continuum frameworks, the corresponding connection between microscopic dynamics and continuum behaviour for helical flows is less developed. Here, we investigate this connection for a dilute gas with time-dependent helical symmetry. Using the objective molecular dynamics framework, we derive a reduced Boltzmann equation that retains radial variations while preserving the classical collision operator under helical symmetry. Formal hydrodynamic expansions then yield the corresponding Euler and Navier--Stokes--Fourier equations. The resulting models describe how cylindrical geometry and axial deformation influence transport, {including viscous transport and heat conduction}, and how deformation generates unequal stresses along and across the axis in competition with collisional relaxation. Comparisons with direct simulation Monte Carlo solutions examine both spatially uniform and radially varying flows. 
        In the cases studied, the first-order fluid corrections capture the leading stress anisotropy and improve the prediction of radial temperature variations, while larger Knudsen numbers expose limitations of the approximation.
	\end{abstract}
	
	\tableofcontents
	\setcounter{tocdepth}{3}
	
	\section{Introduction}
	
	Translational and helical symmetries provide idealized descriptions of flows encountered in nature and engineering {\citep{DuanLiu2021,DierkesOberlack2017}}. Plane shear illustrates the role of translational symmetry, while flows involving rotation about an axis motivate helical descriptions. These macroscopic patterns raise a microscopic question: how can the symmetry of a flow be related to the motion of its constituent particles? Objective molecular dynamics (OMD) provides a framework for describing compatible particle configurations and their evolution \citep{DumitricJames2007,DayalJames2010}. The connection between microscopic dynamics and continuum behaviour has been developed for translation-type structures within the OMD framework {\citep{james2024kinetic}}, whereas the corresponding kinetic reduction and hydrodynamic description for time-dependent helical structures remain less developed. Developing these descriptions is the aim of this work.
	
	The idea connects to a central theme associated with Hilbert's sixth problem {\citep{Hilbert1902}}, which is the mathematical passage from the laws governing individual particles to the continuum equations of gas dynamics. A kinetic
	formulation provides an intermediate description of this passage: a large particle system is represented by a one-particle distribution, and the fluid equations are recovered in a regime in
	which collisions become frequent. The Boltzmann equation is the principal model in this formulation. Its hydrodynamic limits are formally obtained by Hilbert and Chapman--Enskog expansions, with the Knudsen number measuring the relative importance of free transport and
	collisions \citep{Hilbert1902,ChapmanCowling1990}. But why do we need to construct the hydrodynamic systems {at the kinetic level rather than} simply impose symmetry on the Euler or Navier-Stokes equations?
	
	Dierkes and Oberlack developed a time-dependent helical reduction of the incompressible Euler and Navier–Stokes equations that retains two spatial variables {\citep{DierkesOberlack2017}} by imposing the symmetry on the hydrodynamic equations. However, symmetry reduction at the kinetic level constrains both the transport operator (left-hand side of (\ref{eq:boltzmann})) and the collision
	operator, whereas a symmetry imposed directly on the Euler or Navier--Stokes equations would constrain
	only the macroscopic fields and would not determine how the collision invariants, stress
	response, or heat-flux response transform under the same change of variables. Thus, reduction by
	symmetry and passage to the hydrodynamic limit are necessary to keep the
	collision model and constitutive coefficients explicit
	\citep{Caflisch1980,Golse2005,BardosGolseLevermore1993}.
	
	At the kinetic level, let $f=f(t,\y,\vvec)\geq0$ denote the one-particle distribution
	in laboratory position and velocity. In the variables
	$(\y,\vvec)\in\R^3\times\R^3$, the dimensionless equation considered here is
	\begin{equation}
		\label{eq:boltzmann}
		\partial_t f+\boldsymbol{v}\cdot\nabla_{\boldsymbol{y}} f=\frac{1}{\Kn}Q(f,f),
	\end{equation}
	where $\Kn>0$ is the Knudsen number. We use the collision operator
	\begin{equation}
		\label{eq:collision}
		Q(f_i,f_j)(\vvec)=\int_{\R^3}\int_{\Sph^2}
		G\left(|\Delta \boldsymbol{v}|,\frac{\Delta \boldsymbol{v}}{|\Delta \boldsymbol{v}|}\cdot{\boldsymbol d}\right)
		\left[f_i(\vvec')f_j(\vvec_{\ast}')-f_i(\vvec)f_j(\vvec_{\ast})\right]
		\dd{\boldsymbol d}\dd\vvec_{\ast} ,
	\end{equation}
	where $\boldsymbol{d}=(\vvec'-\vvec_{\ast}')/|\Delta \boldsymbol{v}|$ is the unit vector specifying the direction of the post-collision relative velocity, $\Delta \boldsymbol{v}=\vvec-\vvec_{\ast}$ and
	\begin{equation}
		\vvec'=\frac{\vvec+\vvec_{\ast}+|\Delta \boldsymbol{v}|{\boldsymbol d}}{2},\quad
		\vvec_{\ast}'=\frac{\vvec+\vvec_{\ast}-|\Delta \boldsymbol{v}|{\boldsymbol d}}{2}.
	\end{equation}
	The collision invariants are $1$, $\vvec$, and $|\vvec|^2$; hence
	\begin{equation}
		\int_{\R^3}Q(f,f)\phi(\v)\dd\v=0,\quad
		\phi\in\operatorname{span}\{1,\vvec,|\vvec|^2\},
	\end{equation}
	which relates to the local conservation of mass, momentum and kinetic energy.
	The collision operator vanishes at a local Maxwellian under the usual assumptions
	on the collision kernel, and the Boltzmann entropy inequality follows from the
	$H$-theorem {\citep{CercignaniIllnerPulvirenti1994}}.

	The direct numerical treatment of the underlying molecular system is often made possible by
	symmetry. OMD exploits the invariance of interatomic forces under
	Euclidean changes of frame and permutations of identical particles. A group orbit then generates
	the non-simulated particles from a smaller set of simulated particles, while the orbit remains
	an invariant manifold of the molecular dynamics
	\citep{James2006,DumitricJames2007,DayalJames2010}. To connect this reduction with kinetic and continuum descriptions, one must identify how the same
	symmetry acts at these levels. For translation-type OMD
	structures, this question leads to the homoenergetic Boltzmann equation and to corresponding
	reduced hydrodynamic systems \citep{DayalJames2010,james2024kinetic}, where the distribution $f(t,\y,\v)$ is reduced to $g(t,\w)$ with $\w$ being the reduced coupled velocity.
	

	Under time-dependent translational symmetry, James, Nota and Vel\'azquez studied the long-time behaviour of homoenergetic solutions in regimes determined by the relative strength of deformation-induced transport and collisions. They established non-Maxwellian self-similar solutions for Maxwell molecules under suitable conditions \citep{JamesNotaVelazquez2019}, and developed formal asymptotic descriptions of collision-dominated and hyperbolic-dominated regimes \citep{JamesNotaVelazquez2019Collision,JamesNotaVelazquez2020}. In the collision-dominated regime, the leading distribution is Maxwellian with a time-dependent temperature; the hyperbolic-dominated regime exhibits more varied non-equilibrium behaviour. These studies show how deformation and collisional relaxation shape the velocity distribution and its macroscopic moments. We refer the reader to \cite{james2024kinetic} for a systematic summary of the translational case. A special related kinetic reduction, known as the 
	shear affine flow, together with the existence and uniqueness of the solution to the Boltzmann equation, was initially discussed by C. Cercignani \cite{Cercignani1989}.

	For the time-dependent helical structure considered in this work, the relevant isometries combine a rotation about a fixed
	axis with a translation along that axis. In contrast with a purely affine translation reduction,
	the radius $r$ from the helical axis remains a spatial variable. The laboratory distribution is
	consequently represented by a reduced distribution $g(t,r,\w)$, where $\w$ is measured in a
	co-moving helical frame. This raises the question of how the resulting geometric transport terms
	enter the kinetic description and its hydrodynamic approximations.
	
	We derive the corresponding reduced Boltzmann equation, in which the collision operator retains
	its classical form while radial transport, cylindrical curvature and axial dilation enter the
	transport operator. Formal Hilbert and Chapman--Enskog expansions then yield the general reduced
	Euler and Navier--Stokes--Fourier (NSF) systems {\citep{ChapmanCowling1990}}. We focus on the derivation of constitutive laws and do not address the rigorous justification of the hydrodynamic limit; for such a justification for the homoenergetic Boltzmann equation, see \cite{jiang2023hydrodynamic}. The constitutive laws describe how the reduced
	velocity gradients and axial deformation generate viscous stresses, together with heat conduction
	driven by radial temperature gradients. The homogeneous co-moving branch provides a tractable
	special case, with an explicit Maxwell-molecule stress response and a first-Sonine approximation
	for hard spheres.
	The reduction replaces a full three-dimensional kinetic calculation by a representative-cell
	problem while retaining the collision physics. It can therefore make simulations feasible for
	selected helical flows and permits comparisons between collision models. This simplification
	applies on a local symmetry branch where the representative coordinate is well defined. The
	hydrodynamic expansions are formal, and the numerical comparisons assess the resulting fluid
	approximations rather than independently establishing equivalence with full laboratory Boltzmann
	dynamics.

%
{Kinetic calculations can supply stress and heat-flux information to continuum balance laws, as in the multiscale approach of Kessler et al.~\citep{KesslerOranKaplan2010}. Here, we use direct simulation Monte Carlo (DSMC)} {\citep{Bird1994}} {solutions of the reduced kinetic equation to assess the fluid approximations in both radially inhomogeneous and homogeneous settings.} 
{The calculations on a finite radial domain with open boundaries compare} the Euler and NSF predictions for radially varying fields, while the homogeneous
	hard-sphere calculations examine temperature evolution and stress anisotropy over a range of
	Knudsen numbers. These comparisons assess the fluid approximations within the reduced kinetic
	model. A separate reconstruction illustrates the laboratory flow associated with a swirling
	reduced Euler solution.

	The paper is organised as follows. We introduce the Euclidean isometry groups and {the OMD construction first}. We then derive the
	reduced Boltzmann equation in Section 2, including the representative-cell coordinates, the co-moving
	velocity, collision covariance, and the geometric transport terms. In Section 3, we derive the Euler limit, {which introduces}
	the reduced macroscopic fields, the general five-moment system, and the reconstruction of the
	laboratory velocity field. In Section 4, we derive {the Navier--Stokes--Fourier approximation}, which gives the formal Chapman--Enskog
	closure, the explicit Maxwell-molecule response, and the first-Sonine hard-sphere approximation.
	In Section 5, the kinetic-to-fluid comparisons examine stress anisotropy and radial NSF predictions. The final
	section discusses the main results, their scope, and directions for further study.

	\section{Reduced Boltzmann equation under time-dependent helical symmetry}
	We use the framework of objective molecular dynamics (OMD) to reduce the Boltzmann equation under time-dependent helical symmetry, simplifying the simulation and analysis of the resulting flows. James et al. {\citep{DumitricJames2007,DayalJames2010}} showed that molecular dynamics preserves symmetry during nonequilibrium evolution when the rotational part of the defining isometries is time-independent and the translational part varies affinely with time. They investigated translational symmetry, which leads to homoenergetic flows {\citep{DayalJames2010,james2024kinetic}}, and studied the long-time asymptotics of these flows {\citep{JamesNotaVelazquez2019,JamesNotaVelazquez2019Collision,JamesNotaVelazquez2020}}. Here, we consider the helical case to extend the kinetic and hydrodynamic description of symmetry-preserving flows.
	
	\subsection{Objective molecular dynamics and time-dependent helical symmetry}

	{A general symmetry can be described by an isometry group} $\{\mathcal{G}={g_1, g_2,\dots, g_n}\}$ {($n$ may be infinite)} satisfying the axioms of groups, {whose elements $g_i$ are isometries of the form}
	\begin{equation}
		g=(\Qrot\mid\c),\quad\Qrot\in O(3),\quad \c\in\R^3,
		\label{eq:isometry-definition}
	\end{equation}
	with the multiplication and inverse rules
	\begin{equation}
		(\Qrot_1\mid\c_1)(\Qrot_2\mid\c_2)
		=(\Qrot_1\Qrot_2\mid\Qrot_1\c_2+\c_1),
		\quad
		(\Qrot\mid\c)^{-1}=(\Qrot^{T}\mid-\Qrot^{T}\c).
		\label{eq:isometry-algebra}
	\end{equation}
	Under the action 
	\begin{equation}
		g(\y)=\Qrot\y+\c,
	\end{equation}
	the orbit of a point $\y$, i.e. the set
	$\{g(\y):g\in \mathcal{G}\}$, is {a collection of atoms called an {\it objective structure}}. The group action can be applied to a group of atoms. In {\it objective molecular dynamics}, {the trajectories of a finite set of simulated
	atoms are computed using Newtonian dynamics, and the group action generates}
	the remaining atoms. {During the evolution of the dynamical system}, the symmetry is described by the time-dependent isometry group.

	Consider a finite collection of simulated atoms $\y_k(t)$, $k=1,\ldots,M$, and let
	$g_j(t)=(\Qrot_j(t)\mid\c_j(t))$ generate the corresponding atom
	$\y_{j,k}(t)=g_j(t)(\y_k(t))$. Atoms related by the group are taken to have the
	same species and mass. The OMD hypothesis requires the acceleration of every image
	to be the rotated acceleration of its representative:
	\begin{equation}
		\frac{\dd^2}{\dd t^2}g_j(\y_k(t))
		=\Qrot_j\frac{\dd^2\y_k(t)}{\dd t^2}.
		\label{eq:omd-hypothesis}
	\end{equation}
	This hypothesis is natural for a frame-indifferent and permutation-invariant
	interaction potential. In particular, if the force is written formally as
	$m_k\ddot{\y}_{j,k}=-\partial\Phi/\partial\y_{j,k}$, objectivity gives the same
	rotation covariance for the force, while permutation invariance identifies the
	forces on atoms with the same physical attributes. Thus the non-simulated images
	obey the same Newtonian law whenever the OMD acceleration condition holds.
	The time dependence of the isometry is constrained by this condition. Differentiating
	$g_j(t)(\y_k(t))=\Qrot_j(t)\y_k(t)+\c_j(t)$ twice and comparing with the OMD
	acceleration gives, for arbitrary representative trajectories,
	\begin{equation}
		\dot{\Qrot}_j=0,\quad \ddot{\c}_j=0.
		\label{eq:omd-kinematic-condition}
	\end{equation}
	{Consequently, the rotation} is time independent and the translation is affine in time,
	\begin{equation}
		\c_j(t)=\a_j+t\b_j,
		\quad \Qrot_j=\text{constant}.
		\label{eq:omd-translation}
	\end{equation}

	The helical groups satisfying the OMD hypotheses are generated as follows. Let $\{\e_1,\e_2,\e_3\}$ be a right-handed orthonormal basis and $\e=\e_3$ denote the rotation axis of
	$\Qrot_\alpha$ with rotation angle
	$\alpha$. 
	The largest Abelian helical group \citep{Feng2019} is given by
	\begin{equation}
		\mathcal H=\{h^p g^q:p\in\mathbb Z,\ q=0,\ldots,n-1\},
		\label{eq:helical-group-static}
	\end{equation}
	where the generators include a screw displacement with irrational rotation angle
	\begin{equation}
		h=\left(\Qrot_\alpha\,\middle|\,\tau\e+(\I-\Qrot_\alpha)\z\right),
		\quad \frac{\alpha}{2\pi}\notin\mathbb{Q}, \quad \tau\ne0
		\label{eq:static-screw}
	\end{equation}
	and a finite rotation about the same axis
	\begin{equation}
		g=\left(\Qrot_\beta\,\middle|\,(\I-\Qrot_\beta)\z\right),
		\quad \beta=\frac{2\pi}{n},\quad n\in\mathbb{N}^+.
		\label{eq:finite-rotation}
	\end{equation}
	We use the representative-cell coordinates and branch convention stated above; the
	alternative nearest-neighbor reparameterization is not used here
	\citep{Feng2019}.
	
	{In the OMD framework, the constant screw translation is replaced by a
		time-dependent pitch ($\tau = a + b t$). With the origin on the helical axis, set
		\begin{equation}
			\a=(a,0),\quad \b=(b,0),\quad a+bt\ne0
			\label{eq:pitch-condition}
		\end{equation}
		in the time interval under consideration. With $\boldsymbol\xi=(\alpha,\beta)$ and
		$\x=(p,q)$, the action on a representative particle ($\y_{\x,k} = h^p g^q(\y_{k})$) and the corresponding velocity are
		\begin{align}
			\y_{\x,k}(t)&=\Qrot_{\x\cdot\boldsymbol\xi}\y_k(t)
			+(\x\cdot(\a+t\b))\e,\label{eq:helical-position}\\
			\v_{\x,k}(t)&=\Qrot_{\x\cdot\boldsymbol\xi}\dot{\y}_k(t)
			+(\x\cdot\b)\e.\label{eq:helical-velocity}
		\end{align}
		Equations \eqref{eq:helical-position}
		and \eqref{eq:helical-velocity} are the starting point for the kinetic reduction.} In the following discussion, we have one simulated atom $\y_k (t) = \y_0(t)$, and the rest of the atoms are generated by the time-dependent helical group.
	
	\subsection{Invariant ansatz of the Boltzmann equation}
	
	The radius of a point $\y$ from the helical axis is
	\begin{equation}
		r=|(\I-\e\otimes\e)\y|,\quad r>0.
		\label{eq:radius}
	\end{equation}
	For every fixed radius, choose $\y_0=r\e_1$ {as the representative point} and consider the
	continuous representative cell
	\begin{equation}
		\mathcal U(r)=\{h^p g^q(\y_0):0\leq p,q<1\}.
		\label{eq:representative-cell}
	\end{equation}
    {Here, real values of $p$ and $q$ extend the discrete group action
to a continuous family of transformations. The set $\mathcal U(r)$
is a coordinate patch on the cylinder of radius $r$; its images
under the discrete group cover that cylinder. This geometric
extension does not by itself impose continuous invariance on
the distribution function.}
    The point associated with
	$\x=(p,q)$ is
	\begin{equation}
		\y(\x,t)=\Qrot_{\x\cdot\boldsymbol\xi} \y_0
		+(\x\cdot(\a+t\b))\e,
		\label{eq:orbit-coordinate-map}
	\end{equation}
	where $\boldsymbol\xi=(\alpha,\beta)$.
	Conversely, given an arbitrary point $\y$, we may compute the corresponding reference point $\x=(p,q)$ {by inverting the relations}
	\begin{equation}
		\x(\y,t)\cdot(\a+t\b)=\y\cdot\e,
		\quad
		\x(\y,t)\cdot\boldsymbol\xi
		=\operatorname{Arg}(\y\cdot\e_1+i\y\cdot\e_2) \quad (\bmod 2\pi).
		\label{eq:inverse-coordinate}
	\end{equation}
	{This gives $\x(\y,t)=(p(\y,t),q(\y,t))$ in the form}
	\begin{equation}
		p(\y,t)=\frac{\y\cdot\e}{a+bt},
		\quad
		q(\y,t)=\frac{1}{\beta}\operatorname{mod}\left(
		\operatorname{Arg}(\y\cdot\e_1+i\y\cdot\e_2)-p(\y,t)\alpha,2\pi\right).
		\label{eq:inverse-coordinate-explicit}
	\end{equation}
	Equations (\ref{eq:orbit-coordinate-map}) and (\ref{eq:inverse-coordinate-explicit}) {give the correspondence} between the simulated point $\y$ and the reference point $\x$.
	 The formula is local: the angular
	branch, the axis $r=0$, and the possibility that $a+bt=0$ require separate treatment.
	The calculations below are restricted to a domain where these obstructions are
	excluded.
	
	The velocities at the representative point $\y_0$ and the arbitrary point $\y$ are related by the
	active helical isometry. 
    {Under the active helical isometry, a representative particle
with velocity $\dot{\y}_0(t)$ has an image particle with velocity}
	\begin{equation}
		\v(\y,t)=\Qrot_{\x\cdot\boldsymbol\xi}\dot{\y}_0(t)
		+(\x\cdot\b)\e.
		\label{eq:orbit-velocity}
	\end{equation}
	Inspired by the OMD framework, 
    we define the co-moving reduced velocity by
	\begin{equation}
		\w=\Qrot_{-\x(\y,t)\cdot\boldsymbol\xi}\v
		-(\x(\y,t)\cdot\b)\e.
		\label{eq:reduced-velocity}
	\end{equation}
By dotting (\ref{eq:reduced-velocity}) with the absolute coordinates $\e_i$, we observe that the components $w_1$ and $w_2$ are the radial and azimuthal components of the laboratory velocity, respectively, while $w_3=\v\cdot \e-b(\y\cdot \e)/(a+bt)$ is the axial velocity relative to the imposed axial motion.

    {We additionally assume that the distribution is invariant under
this continuous two-parameter action. This is stronger than
invariance under the discrete OMD group or a single continuous
screw motion: the two parameters allow independent rotations
about the axis and axial translations accompanied by the
corresponding velocity shift. We therefore restrict attention
to this OMD-inspired class of distributions, for which the
spatial dependence in the co-moving variables reduces to $r$.
Under this assumption, there is a reduced distribution $g$ such that}
	\begin{equation}
		f(t,\y,\v)= f(t, r\e_1, \w) = g(t,r,\w),
		\label{eq:distribution-reduction}
	\end{equation}
	where
	\begin{align}
		f&:\mathbb R_+\times\mathbb R^3_{\y}\times\mathbb R^3_{\v}\longrightarrow\mathbb R_+,
		\nonumber\\
		g&:\mathbb R_+\times(0,\infty)_r\times\mathbb R^3_{\w}\longrightarrow\mathbb R_+.
		\label{eq:distribution-domains}
	\end{align}
	The reduction is a local statement on the domain where
	$\x(\y,t)$ is a single-valued smooth representative. Here we want to emphasize that if two points are related by
	an integer group operation, they have different laboratory velocities in general,
	but their reduced velocities (i.e. $\w$) coincide after applying the inverse helical motion.
	This is the kinetic counterpart of the distinction between simulated and
	non-simulated particles in OMD. The radius $r$ is retained because the helical orbit
	does not identify points at different distances from the axis.
	
	Recall the laboratory Boltzmann equation \eqref{eq:boltzmann},
	\[
	\partial_t f+\vvec\cdot\nabla_{\y}f=\frac{1}{\Kn}Q(f,f).
	\]
	In the following discussion, we replace the probability distribution $f(t, \y, \v)$ with $g(t, r, \w)$ to derive a reduced Boltzmann equation including the collision term and the convection term. {The spatial dependence is reduced from three coordinates to the radial coordinate, while all three velocity components are retained.}
	
	\subsection{Covariance of the collision operator}
	\label{sec:collision-covariance}
	
	The equivalence of the collision terms can be shown directly by a change of
	variables. Fix $(t,\y)$ and write
	\begin{equation}
		\varphi=\x(\y,t)\cdot\boldsymbol\xi,\quad
		s=\x(\y,t)\cdot\b,\quad
		\vvec=\Qrot_{\varphi}(\w+s\e),
		\label{eq:collision-change-of-variables}
	\end{equation}
	and use the same transformation for the partner velocity,
	$\vvec_{\ast}=\Qrot_{\varphi}(\w_{\ast}+s\e)$. Since this is an affine orthogonal
	transformation, we have
	\begin{equation}
		\dd\vvec_{\ast}=\dd\w_{\ast}, \quad
		\Delta \boldsymbol{w}:=\w-\w_{\ast},\quad
		\Delta \boldsymbol{v}:=\vvec-\vvec_{\ast}=\Qrot_{\varphi}\Delta \boldsymbol{w},\quad |\Delta \boldsymbol{v}|=|\Delta \boldsymbol{w}|.
		\label{eq:collision-relative-change}
	\end{equation}
Since the rotation has unit determinant and the velocity shift is constant at fixed \((t,\boldsymbol y)\), the change of variables preserves the velocity-space volume element, giving \(\mathrm d\boldsymbol v_\ast=\mathrm d\boldsymbol w_\ast\).
For the angular integration, introduce the scattering direction in the rotated frame, $\widetilde{{\boldsymbol d}}=\Qrot_{-\varphi}{\boldsymbol d}$. The surface
	measure on $\Sph^2$ is rotation invariant, and therefore
	\begin{equation}
		\dd{\boldsymbol d}=\dd\widetilde{{\boldsymbol d}},\quad
		\frac{\Delta \boldsymbol{v}}{|\Delta \boldsymbol{v}|}\cdot{\boldsymbol d}
		=\frac{\Delta \boldsymbol{w}}{|\Delta \boldsymbol{w}|}\cdot\widetilde{{\boldsymbol d}}.
		\label{eq:collision-angle-change}
	\end{equation}
	The post-collisional velocities transform in the same way. Indeed, with the
	collision rule of \eqref{eq:collision},
	\begin{align}
		\w'&=\Qrot_{-\varphi}\vvec'-s\e
		=\frac{\w+\w_{\ast}+|\Delta \boldsymbol{w}|\widetilde{{\boldsymbol d}}}{2},\nonumber\\
		\w_{\ast}'&=\Qrot_{-\varphi}\vvec_{\ast}'-s\e
		=\frac{\w+\w_{\ast}-|\Delta \boldsymbol{w}|\widetilde{{\boldsymbol d}}}{2}.
		\label{eq:collision-post-change}
	\end{align}
	Consequently, $f(t,\y,\vvec)=g(t,r,\w)$, and the same relation holds for the
	pre- and post-collisional partner values. Substituting
	\eqref{eq:collision-relative-change}--\eqref{eq:collision-post-change} into the
	collision integral gives the exact identity
	\begin{align}
		Q(f,f)(t,\y,\vvec)
		&=\int_{\R^3}\int_{\Sph^2}
		G\left(|\Delta \boldsymbol{w}|,\frac{\Delta \boldsymbol{w}}{|\Delta \boldsymbol{w}|}\cdot\widetilde{{\boldsymbol d}}\right)
		\left[g(\w')g(\w_{\ast}')-g(\w)g(\w_{\ast})\right]
		\dd\widetilde{{\boldsymbol d}}\dd\w_{\ast}\nonumber\\
		&=Q(g,g)(t,r,\w).
		\label{eq:collision-operator-change}
	\end{align}
	Thus the reduced collision operator has exactly the same mathematical form as the
	laboratory operator, with the velocity variable $\vvec$ replaced by $\w$ and the
	distribution $f(t,\y,\cdot)$ by $g(t,r,\cdot)$. The transformation follows from the unit Jacobian and the rotational covariance of the collision rule. The remaining spatial dependence enters only through $r$.
	
	\subsection{Transport term and the reduced Boltzmann equation}
	\label{sec:reduced-boltzmann-geometry}

	Substituting $f(t,\y,\vvec)=g(t,r,\w)$ and applying the chain rule
	with the time-dependent velocity map \eqref{eq:reduced-velocity} gives
	the transport terms on the left-hand side below. Let
	$\varphi=\operatorname{Arg}(\y\cdot\e_1+i\y\cdot\e_2)$. By direct calculations using
	\begin{equation}
		\nabla_{\y}r=\frac{\y-(\y\cdot\e)\e}{r},\quad
		\nabla_{\y}\varphi=\frac{-\sin\varphi\,\e_1+\cos\varphi\,\e_2}{r},
		\label{eq:radius-angle-gradients}
	\end{equation}
	we obtain
	\begin{equation}
		\partial_t g+(\w\cdot\e_1)\partial_r g
		+\partial_{\w}g\cdot\left[
		-\frac{b}{a+bt}(\w\cdot\e)\e
		-\frac{\w\cdot\e_2}{r}\overline{\boldsymbol W}\w\right]
		=\frac{1}{\Kn}Q(g,g),
		\label{eq:reduced-boltzmann-vector}
	\end{equation}
	where
	$\overline{\boldsymbol W}=-\e_1\otimes\e_2+\e_2\otimes\e_1$
	denotes the rotation generator about $\e=\e_3$.

	In components, Eq.~(\ref{eq:reduced-boltzmann-vector}) becomes
	\begin{equation}
		\partial_t g+w_1\partial_r g
		+\frac{w_2^2}{r}\partial_{w_1}g
		-\frac{w_1w_2}{r}\partial_{w_2}g
		-\frac{bw_3}{a+bt}\partial_{w_3}g
		=\frac{1}{\Kn}Q(g,g).
		\label{eq:reduced-boltzmann-components}
	\end{equation}
	Equivalently, its compact form reads as follows:
	\begin{equation}
		{\cal H}g=\frac{1}{\Kn}Q(g,g),\quad
		\label{eq:reduced-boltzmann-compact}
	\end{equation}
	where
	\begin{equation}
		{\cal H}=\partial_t+w_1\partial_r+\boldsymbol F(\w)\cdot\nabla_{\w},\quad
		\boldsymbol F(\w)=\left(\frac{w_2^2}{r},-\frac{w_1w_2}{r},
		-\frac{bw_3}{a+bt}\right).
		\label{eq:def-H-F}
	\end{equation}
	Here $\cal H$ is the transport operator acting on smooth scalar functions of $(t,r,\w)$ and, at fixed
	$(t,r)$, $\boldsymbol F:\mathbb R^3\to\mathbb R^3$ is the velocity drift.
	The three contributions describe radial transport, cylindrical curvature,
	and axial dilation, respectively. The equation is understood on the smooth
	representative branch with $r>0$ and $a+bt\ne0$. 
	
The reduced Boltzmann equation retains the classical collision operator, while the imposed symmetry enters through the transport terms. The term \(w_1\partial_r g\) describes radial transport, and the two curvature terms account for the changing radial and azimuthal velocity components as particles move through the cylindrical frame. In contrast, the axial term \(-\Gamma w_3\partial_{w_3}g\), where \(\Gamma=b/(a+bt)\), contracts the axial peculiar-velocity distribution during extension and broadens it during compression.  The equation therefore provides a kinetic description of the competition between geometric deformation and collisional relaxation, while retaining radial spatial variations.
	
	\section{The Euler limit}
	In this section, we employ the reduced Boltzmann equation and use the Hilbert expansion to deduce {the reduced Euler equations under} time-dependent helical symmetry. We will discuss how the symmetry enters into the system and how to reconstruct the laboratory flows based on the reduced solution. {A three-dimensional example of the laboratory flow} is presented to illustrate the type of flow. 
	\subsection{General reduced Euler system}
	\label{sec:general-euler}
	{We use $\braket{a,b}$ to denote the unweighted velocity-space pairing, defined for real-valued functions by}
	{
	\[
	\braket{a,b}=\int_{\mathbb R^3}a(\w)b(\w)\,\dd\w.
	\]
	}
	{If one argument is vector- or tensor-valued, the integral is taken componentwise.}
	For the reduced distribution, define the moments
	\begin{equation}
		\rho=\braket{g,1},\quad
		\boldsymbol u=\frac{1}{\rho}\braket{g,\w},\quad
		\theta=\frac{1}{3\rho}\braket{g,|\w-\boldsymbol u|^2}.
		\label{eq:reduced-fields}
	\end{equation}
	The vector $\boldsymbol u$ is the reduced bulk velocity, not the laboratory
	velocity. In particular, the restriction $\boldsymbol u=0$ selects a co-moving
	state with no residual drift after the helical motion has been removed; it does not
	assert that the physical laboratory velocity vanishes.
	
	{The change of variables in Section~\ref{sec:collision-covariance}} does not alter the velocity-space algebra of the
	collision operator. Consequently, the standard properties of $Q(f,f)$ can be used
	directly for $Q(g,g)$. For every sufficiently integrable reduced distribution, {the collision-invariant moments satisfy} {\citep{CercignaniIllnerPulvirenti1994}} 
	\begin{equation}
		\braket{Q(g,g),\phi(\w)}=0,
		\quad \phi(\w)\in\operatorname{span}\{1,\w,|\w|^2\}.
		\label{eq:reduced-collision-invariants}
	\end{equation}
	The equality case of the entropy production is the local Maxwellian family. Thus,
	for some $\rho>0$, reduced bulk velocity $\boldsymbol u$, and temperature
	$\theta>0$,
	\begin{equation}
		g=\rho M_{\boldsymbol u,\theta},
		\quad
		M_{\boldsymbol u,\theta}(\w)=
		(2\pi\theta)^{-3/2}
		\exp\left(-\frac{|\w-\boldsymbol u|^2}{2\theta}\right),
		\label{eq:reduced-maxwellian-family}
	\end{equation}
	and the $H$-theorem takes the same form,
	\begin{equation}
		\braket{\log g,Q(g,g)}\leq0.
		\label{eq:reduced-H-theorem}
	\end{equation}
	These statements are mathematical properties of the reduced velocity equation. The
	velocity map is a rotation followed by a translation with unit Jacobian, so the
	laboratory and reduced densities and centered temperatures agree; mean velocities and
	stress components are related by the active rotation.
	
	The Euler closure follows from the Hilbert expansion
	\begin{equation}
		g=\sum_{k=0}^{\infty}\Kn^k g_k.
		\label{eq:hilbert-expansion}
	\end{equation}
	Substitution into ${\cal H} g=\Kn^{-1}Q(g,g)$ and comparison of equal powers of            
	$\Kn$ give
	\begin{align}
		\Kn^{-1}:\quad &Q(g_0,g_0)=0, \nonumber\\
		\Kn^{0}:\quad &{\cal H} g_0=Q(g_0,g_1)+Q(g_1,g_0),\nonumber\\
		\Kn^{n-1}:\quad &{\cal H} g_{n-1}=\sum_{k=0}^{n}Q(g_k,g_{n-k}),
		\quad n=2,3,\ldots .\label{eq:hilbert-order-n}
	\end{align}
	Here ${\cal H}$ is the reduced transport operator defined in \eqref{eq:reduced-boltzmann-compact}.
	Since the reduced collision operator has the same mathematical form as the
	classical one, we follow the moment argument in Bobylev
	\citep[Section~7.2]{Bobylev2020}. The collision-invariant
	moments of $Q(g_0,g_1)+Q(g_1,g_0)$ vanish. Thus, taking the five moments (\ref{eq:reduced-collision-invariants})
	and substituting the local Maxwellian
	$g_0=\rho M_{\boldsymbol u,\theta}$ {into the zeroth-order equation in \eqref{eq:hilbert-order-n}} give the reduced Euler equations.
	
	We assume a smooth representative branch with $r>0$,
	$a+bt\ne0$, and sufficient velocity decay for integration by parts.
	Set
	\begin{equation}
		D_t=\partial_t+u_1\partial_r,\quad p=\rho\theta.
		\label{eq:reduced-material-derivative}
	\end{equation}
	For a scalar field depending only on $(t,r)$, $D_t$ is the material
	derivative: it measures the rate of change following the mean flow,
	whose radial velocity is $u_1$. For the velocity components in the
	rotating cylindrical basis, the geometric acceleration terms must also
	be included. By direct calculations, the resulting system is
	\begin{equation}
		\left\{\begin{aligned}
			&D_t\rho+\rho\left(\partial_r u_1+\frac{u_1}{r}+\frac{b}{a+bt}\right)=0,\\
			&D_tu_1+\frac{1}{\rho}\partial_r p-\frac{u_2^2}{r}=0,\\
			&D_tu_2+\frac{u_1u_2}{r}=0,\\
			&D_tu_3+\frac{b}{a+bt}u_3=0.
		\end{aligned}\right.
		\label{eq:general-euler-primitive}
	\end{equation}
	{with the temperature equation} 
	\begin{equation}
		D_t\theta+\frac23\theta\left(\partial_r u_1+\frac{u_1}{r}
		+\frac{b}{a+bt}\right)=0.
		\label{eq:general-euler-temperature}
	\end{equation}
	The Euler system can be written in a compact form as follows.
	
	\paragraph{Compact form of the reduced Euler system.}
	Using $\nabla_r f=\e_1\partial_r f$ and
	$\operatorname{div}_r\boldsymbol h=r^{-1}\partial_r(rh_1)$,
	and writing $\Gamma=b/(a+bt)$ and
	$\Theta=\operatorname{div}_r\boldsymbol u+\Gamma$, we obtain
	\begin{equation}
		\boxed{\begin{aligned}
				D_t\rho+\rho\Theta&=0,\\
				\rho\bigl(D_t\boldsymbol u-\boldsymbol F(\boldsymbol u)\bigr)
				+\nabla_r p&=0,\\
				\frac32\rho D_t\theta+p\Theta&=0,
				\quad p=\rho\theta.
		\end{aligned}}
		\label{eq:compact-euler}
	\end{equation}
	Here $\boldsymbol F$ is exactly the velocity drift in
	\eqref{eq:def-H-F}, evaluated at $\boldsymbol u$.
	{The curvature acceleration and the axial dilation are therefore retained.}
	$\Gamma$ is the axial strain rate of the imposed motion:
	positive and negative values describe axial extension and compression,
	respectively. The quantity $\Theta$ is the divergence of the laboratory
	bulk velocity reconstructed below, hence the local rate of volume expansion
	per unit volume. It includes both radial expansion and axial stretching;
	$\Theta>0$ gives local dilution and adiabatic cooling.

	\subsection{Numerical validation for the general Euler system}
	\label{sec:general-euler-validation}
	
	We compare the general reduced Euler system with a hard-sphere DSMC solution
	of the reduced Boltzmann equation on $0.5\leq r\leq2.5$, $0\leq t\leq0.5$.
	All variables in this test are dimensionless. We set $a=1$, $b=1/2$,
	$d=1/2$, and $\Kn=0.025$, with the collision kernel
	$G=|\w-\w_{\ast}|d^2/4$ and the ordinary area measure on $\Sph^2$.
	The kinetic initial state is a local Maxwellian with moments
	\begin{equation}
		\left\{\begin{aligned}
			&\rho(0,r)=1+0.04\sin\bigl(\pi(r-0.5)\bigr),\\
			&u_1(0,r)=0.03\sin\bigl(\pi(r-0.5)/2\bigr),\\
			&u_2(0,r)=0,\quad u_3(0,r)=0.08,\quad\theta(0,r)=1.
		\end{aligned}\right.
		\label{eq:general-euler-validation-data}
	\end{equation}
	
{Open boundaries are imposed at both radial ends. The Euler solver uses zero-gradient ghost states.
In DSMC, outgoing particles leave the domain, while incoming particles are sampled from the flux-weighted Maxwellian constructed from the adjacent interior cell's density, mean velocity, and temperature.
This permits both inflow and outflow without a reflecting wall and approximates the gas outside the finite computational domain.
The nonuniform density and radial velocity make this a spatially inhomogeneous test.
Although the initial temperature is uniform, the nonuniform expansion rate in \eqref{eq:general-euler-temperature} generates radial temperature variations.}
{The sensitivity of rarefied-flow predictions to kinetic boundary conditions is illustrated by the gas--surface comparisons of Wu and Struchtrup \citep{WuStruchtrup2017}, but their wall models differ from the open-reservoir approximation used here.}
	
	For this comparison, the Euler equations are discretized by a finite-volume method on 640 radial
	cells, with a Rusanov flux and a third-order strong-stability-preserving Runge--Kutta time
	integrator with CFL number $0.45$ and maximum step $0.00025$. The cell states are averages with
	respect to $r\,\dd r$. Advancing $(a+bt)\bm{\mathcal U}$, where $\bm{\mathcal U}$ denotes the
	five conservative moments, accounts for the pitch-dependent density source. The geometric flux
	and source quadratures preserve a constant pressure at rest.
	
	The DSMC calculation uses 640 collision cells, a time step $\Delta t=0.00025$, and approximately
	1200 particles per collision cell on average. Eight independent replicas are initialized with
	paired Gaussian velocities, rescaled to match the prescribed cell temperature. Positions are
	stratified with respect to $r\,\dd r$. 
{Equal particle weights represent $h=(a+bt)r g$; the mass within the computational domain changes by the difference between incoming and outgoing particle fluxes.} Each step consists of a half
	transport step, cellwise collisions, and a second half transport step. Transport follows the
	reduced characteristics, 
{including radial transport through the open ends and axial velocity rescaling.} Hard-sphere
	collisions use majorant-based acceptance sampling; each accepted pair conserves momentum and
	kinetic energy. The axial drift changes the reduced kinetic energy during transport.
	
	Temperature profiles are reported at intervals of $0.1$. For comparison,
	both numerical solutions are aggregated to 20 common radial observation
	intervals using the conservative moments and the measure $r\,\dd r$;
	temperature is reconstructed after aggregation. The DSMC mean and its standard
	error are then computed across the eight replicas. The relative temperature
	error at a fixed time is
	\begin{equation}
		E_\theta(t)=\left[
		\frac{\sum_{j=1}^{20}\bigl(\overline{\theta}_{\mathrm D,j}(t)
			-\theta_{\mathrm E,j}(t)\bigr)^2}
		{\sum_{j=1}^{20}\theta_{\mathrm E,j}(t)^2}
		\right]^{1/2},
		\label{eq:general-euler-validation-error}
	\end{equation}
	where the bar denotes the replica mean. This normalization uses temperature
	itself, rather than the amplitude of its radial variation.

\begin{figure}[H]
		\centering
		\includegraphics[width=0.98\linewidth]{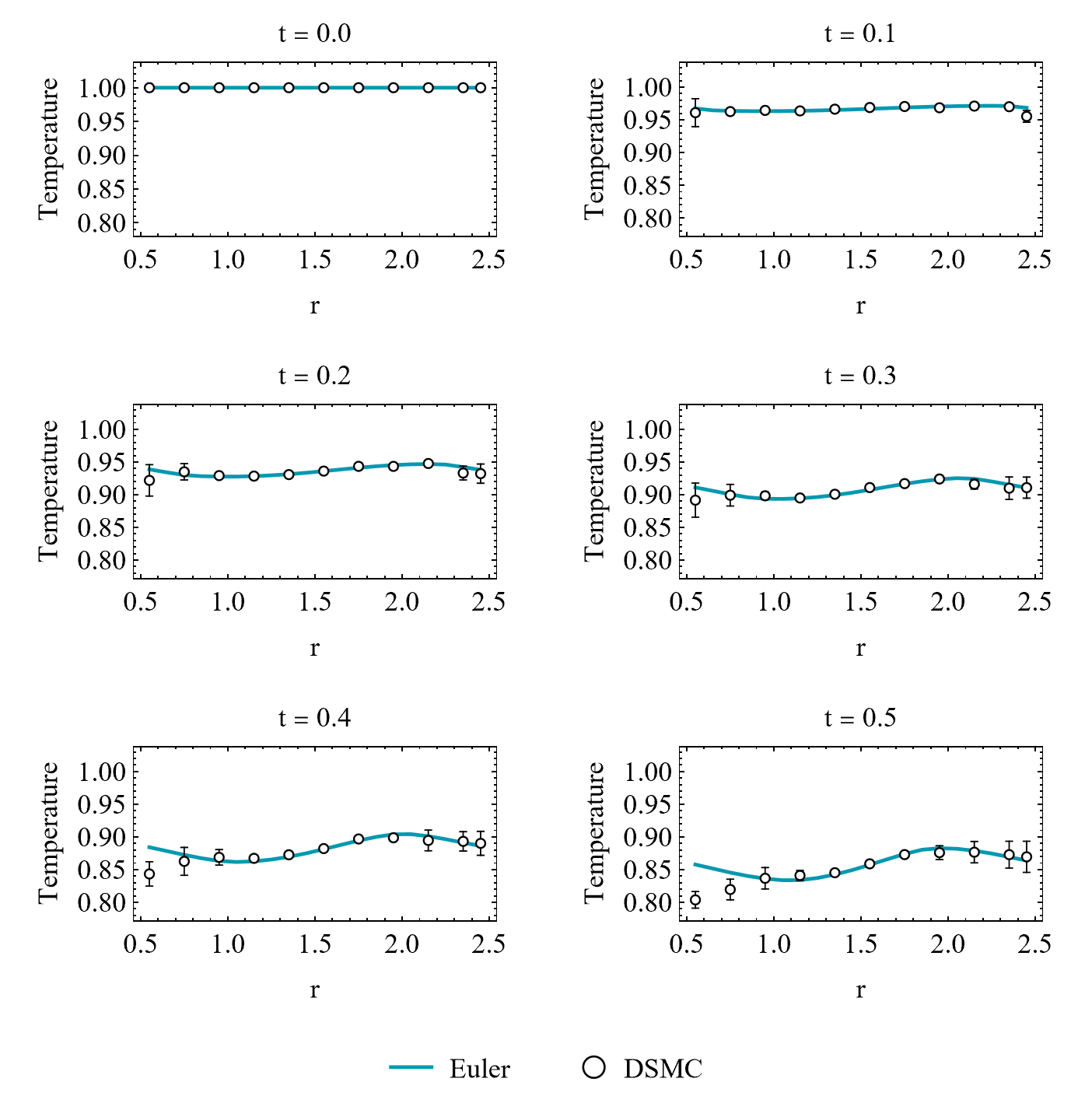}
\caption{{Radial temperature profiles at six times with open radial boundaries: Euler and DSMC. Error bars indicate one standard error across eight replicas.}}
		\label{fig:radial-target640}
	\end{figure}

{At the final saved time, the relative temperature RMS error is $1.8763\%$, and the corresponding density RMS error is $2.9243\%$.
The global temperature reconstructed from the total mass, momentum, and energy is $0.86416$ for DSMC and $0.86275$ for Euler, with a relative bias of $0.1645\%$; the DSMC standard error is $0.01006$.
Figure~\ref{fig:radial-target640} shows the radial temperature profiles at six times.
The largest discrepancies occur near the inner boundary, where the replica variation is also substantial.}
	
\subsection{Reconstruction of the laboratory flow}    

	The reduced velocity $\boldsymbol u(t,r)$ is the mean in the co-moving helical
	frame, not the laboratory velocity. Its laboratory reconstruction follows from the
	inverse velocity map:
	\begin{equation}
		\boldsymbol U(t,\y)=\Qrot_{\varphi}\boldsymbol u(t,r)
		+\Gamma(t)z\e_3,\quad
		\Gamma(t)=\frac{b}{a+bt},\quad z=\y\cdot\e_3.
		\label{eq:lab-flow-reconstruction}
	\end{equation}
	Here $\Qrot_\varphi$ is the active rotation from reduced cylindrical components
	to the laboratory basis, and \(\varphi\) is the azimuthal angle of the laboratory position \(\boldsymbol y\).  The density and centered temperature are unchanged by the
	velocity-space rotation and translation. A streamline at a fixed time solves
	$\dd\y/\dd s=\boldsymbol U(t,\y)$ with $t$ held fixed. Here \(s\) is a parameter tracing the streamline.
    \begin{figure}[!h]
		\centering
		\includegraphics[width=.9\linewidth]{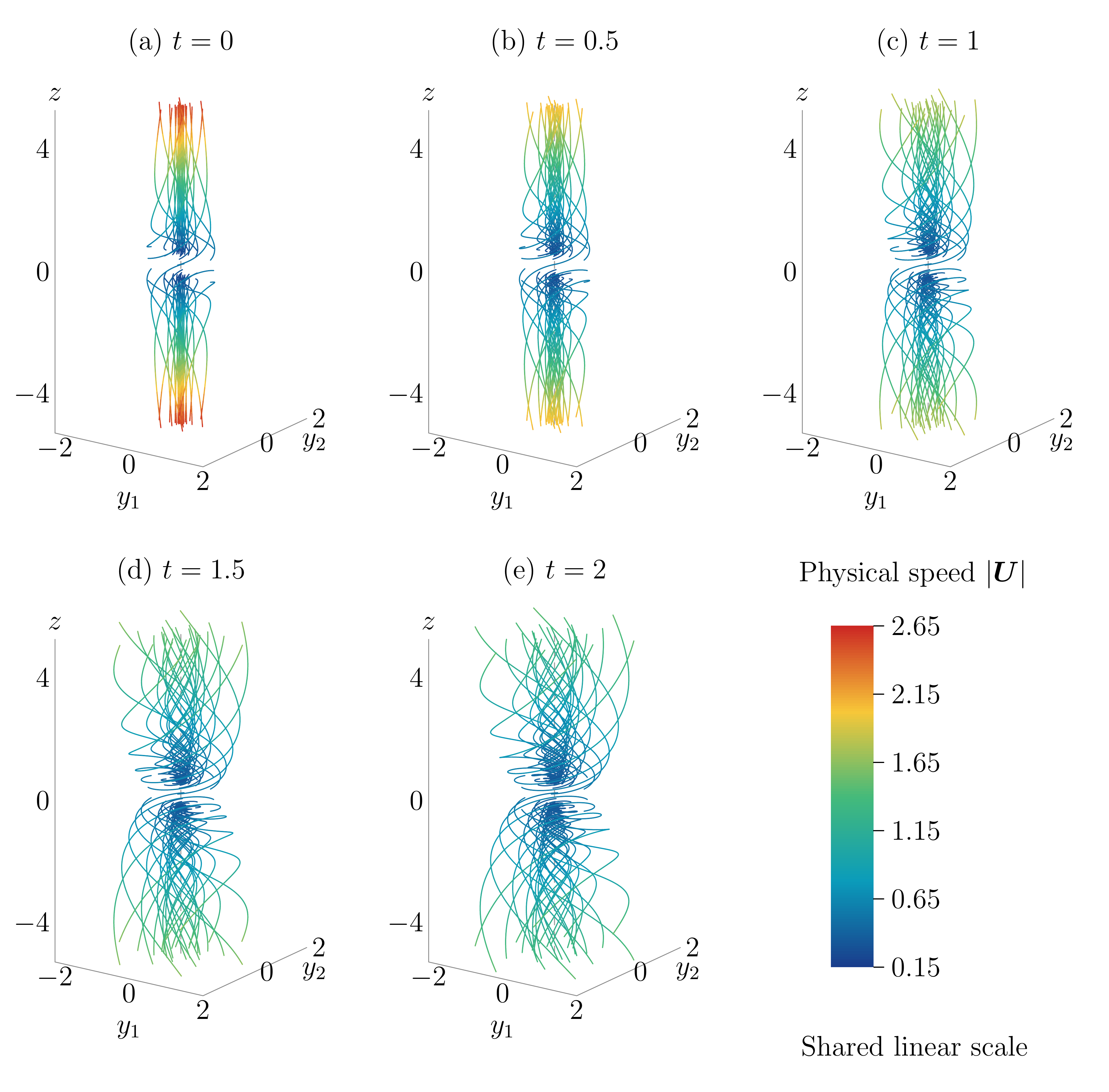}
		\caption{Instantaneous laboratory-flow streamlines at
			$t=0,0.5,1,1.5,2$. Color indicates physical speed $|\boldsymbol U|$
			on a shared linear scale.}
		\label{fig:near-axis-streamlines}
	\end{figure}

	For the following example, all variables are dimensionless. We use $a=1$, $b=0.5$, $\Omega=0.65$,
	$R_0=3$, and $\theta_0=1.5$, with initial data $u_1(0,r)=-0.08r\exp(-r^2/R_0^2)$,
	$u_2(0,r)=\Omega r\exp(-r^2/R_0^2)$, $u_3(0,r)=0$, $\theta(0,r)=\theta_0$, and
	$\rho(0,r)=\exp[-\Omega^2R_0^2\exp(-2r^2/R_0^2)/(4\theta_0)]$. The initial pressure gradient
	balances the centrifugal term, while the inward radial velocity initiates the evolution. The
	laboratory axial velocity is initially $U_z=0.5z$.
	
	The Euler solution uses 320 radial finite-volume cells on $0\leq r\leq20$, with a Rusanov flux,
	third-order strong-stability-preserving Runge--Kutta integration, CFL number $0.4$, and maximum
	time step $0.004$. At the axis, scalar fields and $u_3$ are extended evenly and $u_1,u_2$ oddly.
	The outer boundary uses zero-gradient ghost states, allowing inflow and outflow, and it approximates
	an unbounded radial domain.
	
	Figure~\ref{fig:near-axis-streamlines} shows the near-axis reconstruction at $t=0,0.5,1,1.5,2$.
	Sixty streamlines start at radii $0.05,0.1,0.2,0.4,0.8$, with six equally spaced azimuths and
	$z=\pm0.36$, and are followed towards $|z|=5.2$. All panels use the same view and windows
	$[-2,2]^2$ and $-5.25\leq z\leq5.25$, with unequal transverse and axial display scales.
	Near-axis interpolation preserves the regular parity conditions but adds no spatial resolution.
	This figure illustrates the geometry of the Euler flow.
	
	\subsection{Special homogeneous Euler branch}
	
	The Euler equations simplify further for zero reduced bulk velocity
	and radial independence:
	\begin{equation}
		\boldsymbol u=0,
		\quad \partial_r\rho=0,
		\quad \partial_r\theta=0.
		\label{eq:homogeneous-zero-mean-assumptions}
	\end{equation}
	This special state describes axial extension or compression without
	residual motion in the co-moving frame. The mass and temperature equations
	reduce to
	\begin{equation}
		\left\{
		\begin{aligned}
			&\rho_t+\frac{b}{a+bt}\rho=0,\\
			&\theta_t+\frac{2b}{3(a+bt)}\theta=0.
		\end{aligned}\right.
		\label{eq:reduced-euler}
	\end{equation}
	
	The two homogeneous equations can also be read directly as scaling laws. On an
	interval where $a+bt$ does not change sign,
	\begin{equation}
		\rho(t)=\rho_0\frac{a+bt_0}{a+bt},
		\quad
		\theta_{\rm E}(t)=\theta_0
		\left(\frac{a+bt_0}{a+bt}\right)^{2/3}.
		\label{eq:euler-scaling-laws}
	\end{equation}
	Here $\rho_0$ and $\theta_0$ are the values at $t=t_0$.
	This analytic solution of the homogeneous Euler system describes dilution
	and adiabatic cooling during axial extension, or concentration and heating
	during compression. With $\boldsymbol u=0$ in
	\eqref{eq:lab-flow-reconstruction}, the laboratory flow is purely axial
	and has no mean swirl. It is a homoenergetic affine flow {in the sense of Cercignani \citep{Cercignani1989}}.
	
	\section{The Navier--Stokes--Fourier approximation}
	The Euler system derived in the preceding section describes the leading-order dynamics but does not account for viscous stresses or heat conduction. To capture these effects at small but finite Knudsen number, we derive the first-order Chapman--Enskog correction to the reduced distribution. This correction yields the constitutive relations and the general reduced Navier--Stokes--Fourier system, retaining the contributions of cylindrical geometry and axial deformation. We then examine a homogeneous branch to obtain analytical reference solutions and compare the hard-sphere predictions with DSMC results, using a first-Sonine approximation for the transport coefficients.
	\subsection{Chapman--Enskog correction}
	
	The Euler approximation retains only the local Maxwellian. The first
	kinetic correction is written as
	\begin{equation}
		g=\rho M+\Kn G+O(\Kn^2).
		\label{eq:chapman-ansatz}
	\end{equation}
	Let $\cvar=\w-\boldsymbol u$ be the peculiar velocity.  Expanding the centered
	moments gives
	\begin{equation}
		\begin{aligned}
			\braket{G,1}&=0,\\
			\braket{G,\cvar}
			&=\braket{G,\w}-\boldsymbol u\braket{G,1}=0,\\
			\braket{G,|\cvar|^2}
			&=\braket{G,|\w|^2}-2\boldsymbol u\cdot\braket{G,\w}
			+|\boldsymbol u|^2\braket{G,1}=0.
		\end{aligned}
		\label{eq:chapman-centered-moments}
	\end{equation}
	Thus the correction also has zero centered mass, momentum, and energy
	moments. The normalized Maxwellian is
	\begin{equation}
		M(\cvar)=(2\pi\theta)^{-3/2}
		\exp\left(-\frac{|\cvar|^2}{2\theta}\right).
		\label{eq:maxwellian}
	\end{equation}
	
	Set $h=G/(\rho M)$ and define the linearized collision operator ${\cal L}$ by
	\begin{equation}
		{\cal L}h=M^{-1}[Q(Mh,M)+Q(M,Mh)].
		\label{eq:linearized-operator}
	\end{equation}
We adopt the non-positive convention for \(\cal L\), so that \(\langle h,{\cal L}h\rangle_M\leq0\), reflecting the dissipative effect of collisions. Substitution into the reduced
	Boltzmann equation and collection of the terms of order $\Kn^0$ formally give
	\begin{equation}
		{\cal L}h=\frac{1}{\rho^2M}{\cal H}(\rho M).
		\label{eq:first-chapman-equation}
	\end{equation}
	The factor $\rho^2$ comes from the two Maxwellian factors in the
	bilinear collision term.
	
	For the collision kernels considered here, the standard nullspace
	property is $\ker L=\operatorname{span}\{1,c_1,c_2,c_3,|\cvar|^2\}$
	\citep{CercignaniIllnerPulvirenti1994,ChapmanCowling1990}.
	Thus the source must be orthogonal to these collision invariants.
	Let $\Pi$ be the orthogonal projection onto $\ker {\cal L}$ in the
	Maxwellian inner product. The moment conditions in
	\eqref{eq:chapman-centered-moments} select the solution satisfying
	\begin{equation}
		\Pi h=0.
		\label{eq:chapman-projection-gauge}
	\end{equation}
	The equation for the nonequilibrium correction therefore takes the form
	\begin{equation}
		{\cal L}h=S^\perp,\quad
		S^\perp=(\mathrm{I}-\Pi)\left[\frac{1}{\rho^2M}{\cal H}(\rho M)\right].
		\label{eq:projected-chapman-source}
	\end{equation}
	The removed moments give the Euler solvability conditions; when these
	hold, the unprojected source is already orthogonal to $\ker {\cal L}$.
	The inverse of ${\cal L}$ is consequently understood on $(\ker {\cal L})^\perp$.
	
	For a moving Maxwellian, we seek the projected source in the two
	Burnett-function sectors familiar from the classical Chapman--Enskog
	expansion \citep{ChapmanCowling1990}:
	\begin{equation}
		\begin{aligned}
			S^\perp
			&= A_{ij}(t,r)\left(c_ic_j-\frac{|\cvar|^2}{3}\delta_{ij}\right)
			+B_i(t,r)c_i\left(\frac{|\cvar|^2}{2\theta}-\frac{5}{2}\right) \\ 
			&=A_{ij}(t,r) Q_{ij}(\boldsymbol{c}) + B_i(t,r) H_i(\boldsymbol{c}).
			\label{eq:general-chapman-source}
		\end{aligned}
	\end{equation}
	The traceless quadratic tensor describes the stress response, while the
	cubic vector describes heat transport. Both are orthogonal to the
	collision invariants. {The following calculation gives the coefficients}
	from the velocity gradients, radial geometry, axial strain rate, and
	temperature gradient.

	To determine the coefficients $A_{ij}$ and $B_i$
	in the projected source $S^\perp$, we introduce the
	deformation tensor and its traceless part:
	\begin{align}
		&\boldsymbol{D}=\begin{pmatrix}
			u_{1,r}&(u_{2,r}-u_2/r)/2&u_{3,r}/2\\
			(u_{2,r}-u_2/r)/2&u_1/r&0\\
			u_{3,r}/2&0&\Gamma
		\end{pmatrix},\nonumber\\
		&\bar{\boldsymbol{D}}=\boldsymbol{D}-\frac{\Theta}{3}\I,
		\quad \Theta=\operatorname{tr}\boldsymbol D.
		\label{eq:general-nsf-deformation}
	\end{align}
	The quantities $\Gamma$ and $\Theta$ are the same as in the {compact Euler system~\eqref{eq:compact-euler}.
	Substituting} the
	local Maxwellian into \eqref{eq:projected-chapman-source} and projecting
	out the collision invariants gives
	\begin{equation}
		{S^\perp=
			\frac{\bar{D}_{ij}}{\rho\theta} Q_{ij}
			+\frac{\theta_r}{\rho\theta} H_1.}
		\label{eq:general-nsf-explicit-source}
	\end{equation}

	\subsection{Constitutive relation for the general reduced NSF system}
	\label{sec:general-nsf-system}
	
	The projected source determines the first-order correction through
	$G=\rho M {\cal L}^{-1}S^\perp$, with the inverse taken on $(\ker {\cal L})^\perp$.
	Only its stress and heat-flux moments are needed to close the macroscopic
	balances. Rotational invariance of the collision operator makes their
	response independent of direction, yielding scalar viscosity and thermal
	conductivity coefficients.
	
	Define the centered pressure tensor and heat flux by
	\begin{equation}
		\boldsymbol{P}=\int_{\mathbb R^3}\cvar\otimes\cvar\,g\,\dd\w=p\I+\boldsymbol{\tau},
		\quad \boldsymbol{q}=\frac12\int_{\mathbb R^3}|\cvar|^2\cvar\,g\,\dd\w,
		\quad p=\rho\theta.
		\label{eq:general-nsf-moments}
	\end{equation}
	Here $\boldsymbol{\tau}=\boldsymbol{P}-p\I$ denotes the pressure deviation and includes the factor $\Kn$, {and is the negative of the viscous Cauchy stress.} The resulting first-order constitutive laws are
	\begin{equation}
		{\boldsymbol{P}=p\I-2\Kn\mu(\theta)\bar{\boldsymbol{D}},\quad
			\boldsymbol{q}=-\Kn\kappa(\theta)\theta_r\boldsymbol e_1,}
		\label{eq:general-nsf-laws}
	\end{equation}
	where the coefficients do not include $\Kn$ and are defined by
	\begin{align}
		\mu(\theta)&=-\frac{1}{10\theta}
		\sum_{i,j}\braket{ Q_{ij},{\cal L}^{-1} Q_{ij}}_M
		=-\frac{3}{4\theta}\braket{ Q_{33},{\cal L}^{-1} Q_{33}}_M,\label{eq:general-nsf-mu}\\
		\kappa(\theta)&=-\frac13\sum_i\braket{ H_i,{\cal L}^{-1} H_i}_M
		=-\braket{ H_1,{\cal L}^{-1} H_1}_M.\label{eq:general-nsf-kappa}
	\end{align}
	The non-positive sign convention for $\cal L$ makes these coefficients positive.
	The pressure and heat-flux laws approximate the kinetic moments through first
	order. {The subscript $M$ denotes the Maxwellian-weighted inner product,}
	{
	\[
	\braket{a,b}_M=\int_{\mathbb R^3}M(\cvar)a(\cvar)b(\cvar)\,\dd\cvar
	=\braket{Ma,b}.
	\]
	}
	
	Substitution into the exact reduced moment balances closes five equations
	\begin{equation}
		\left \{    
		\begin{aligned}
			&D_t\rho+\rho\Theta=0,\\
			&\rho D_tu_1=\frac{\rho u_2^2}{r}-p_r
			-\partial_r\tau_{11}-\frac{\tau_{11}-\tau_{22}}{r},\\
			&\rho D_tu_2=-\frac{\rho u_1u_2}{r}
			-\partial_r\tau_{12}-\frac{2\tau_{12}}{r},\\
			&\rho D_tu_3=-\frac{b}{a+bt}\rho u_3
			-\partial_r\tau_{13}-\frac{\tau_{13}}{r},\\
			&\frac32\rho D_t\theta+p\Theta
			=2\Kn\mu \bar{\boldsymbol{D}}:\bar{\boldsymbol{D}}+\frac{\Kn}{r}\partial_r(r\kappa\theta_r),
		\end{aligned} \right. \label{eq:nsf-temperature}
	\end{equation}
	{with the explicit stress components}
	\begin{align}
		\tau_{11}&=-2\Kn\mu(u_{1,r}-\Theta/3),&
		\tau_{22}&=-2\Kn\mu(u_1/r-\Theta/3),\nonumber\\
		\tau_{33}&=-2\Kn\mu\left(\frac{b}{a+bt}-\frac\Theta3\right),&
		\tau_{12}&=-\Kn\mu(u_{2,r}-u_2/r),\nonumber\\
		\tau_{13}&=-\Kn\mu u_{3,r},&\tau_{23}&=0.
		\label{eq:nsf-stress-components}
	\end{align}
	The diagonal entries of $\boldsymbol\tau$ describe departures of the
	normal pressures from $p$, while its off-diagonal entries describe shear
	momentum transport. The term $2\Kn\mu\bar{\boldsymbol D}:\bar{\boldsymbol D}$
	converts mean-flow kinetic energy into heat; the remaining term redistributes
	heat by conduction. Since temperature depends only on $(t,r)$, the heat
	flux is radial. Both transport coefficients remain inside the spatial
	derivatives.
	
	\paragraph{Compact form of the reduced NSF system.}
	For a symmetric tensor field $\boldsymbol T$, define the reduced tensor divergence as
	\begin{equation}
		\operatorname{Div}_r\boldsymbol T=\partial_r(\boldsymbol T\e_1)
		+\frac1r\begin{pmatrix}T_{11}-T_{22}\\2T_{12}\\T_{13}\end{pmatrix}.
		\label{eq:compact-operators}
	\end{equation}
	Together with the scalar gradient and vector divergence used in
	Section~\ref{sec:general-euler}, this gives
	\begin{equation}
		\boxed{\begin{aligned}
				D_t\rho+\rho\Theta&=0,\\
				\rho\bigl(D_t\boldsymbol u-\boldsymbol F(\boldsymbol u)\bigr)
				+\nabla_r p&=2\Kn\operatorname{Div}_r(\mu \bar{\boldsymbol{D}}),\\
				\frac32\rho D_t\theta+p\Theta
				&=2\Kn\mu \bar{\boldsymbol{D}}:\bar{\boldsymbol{D}}
				+\Kn\operatorname{div}_r(\kappa\nabla_r\theta).
		\end{aligned}}
		\label{eq:compact-nsf}
	\end{equation}
	Setting $\Kn=0$ recovers the compact Euler {system~\eqref{eq:compact-euler}.}
	
	\subsection{Benchmarks of homogeneous analytical NSF branches for hard-sphere molecules}
	On the homogeneous zero-mean branch \eqref{eq:homogeneous-zero-mean-assumptions}, the heat flux vanishes and the
	general NSF system reduces to
	\begin{equation}
		\rho_t+\frac{b}{a+bt}\rho=0,\quad
		\theta_t+\frac{2b}{3(a+bt)}\theta
		-\frac{8\Kn\mu}{9\rho}\left(\frac{b}{a+bt}\right)^2=0.
		\label{eq:general-nsf-homogeneous}
	\end{equation}
	The collision model determines $\mu$ and hence the temperature evolution.

	For hard spheres, the collision kernel is
	\begin{equation}
		G_{\mathrm{HS}}(|\Delta \boldsymbol{v}|,\mu)=\frac{d^2}{4}|\Delta \boldsymbol{v}|,
		\label{eq:hard-sphere-kernel}
	\end{equation}
	which is the standard rigid-sphere model in the stated normalization
	with standard surface measure on $\Sph^2$
	\citep{Cercignani1972,ChapmanCowling1990}. The degree-two tensor sector is
	preserved by the rotationally invariant collision operator. The auxiliary solution
	$h=L^{-1}Q_{33}$ therefore has the form
	\begin{equation}
		h(\cvar)=\chi(|\cvar|)Q_{33}(\cvar).
		\label{eq:hard-sphere-spectral}
	\end{equation}
	The first Sonine approximation retains only the constant term in the radial
	factor:
	\begin{equation}
		\chi(|\cvar|)\approx a_0,\quad h^{(0)}=a_0Q_{33}.
		\label{eq:first-sonine-ansatz}
	\end{equation}
	Requiring the residual ${\cal L}h^{(0)}-Q_{33}$ to be orthogonal to
	$Q_{33}$ in the Maxwellian inner product gives the Galerkin coefficient
	\begin{equation}
		a_0=\frac{\braket{Q_{33},Q_{33}}_M}
		{\braket{Q_{33},LQ_{33}}_M},
		\label{eq:first-sonine-coefficient}
	\end{equation}
	where $\braket{g_1,g_2}_M=\braket{Mg_1,g_2}$.
	Substituting this approximation and the analogous lowest Sonine approximation
	for ${\cal L}^{-1}H_1$ into the definitions
	\eqref{eq:general-nsf-mu}--\eqref{eq:general-nsf-kappa} gives
	\begin{equation}
		\mu^{(0)}=\frac{5}{16d^2}\sqrt{\frac\theta\pi},\quad
		\kappa^{(0)}=\frac{75}{64d^2}\sqrt{\frac\theta\pi}.
		\label{eq:hard-sphere-general-coefficients}
	\end{equation}
	These are first-Sonine approximations to the hard-sphere transport coefficients, and
	the contribution of higher Sonine terms is not estimated here.
	
	With this normalization, {the homogeneous hard-sphere reduced-NS system is}
	\begin{equation}
		\left\{
		\begin{aligned}
			&\rho_t+\frac{b}{a+bt}\rho=0,\\
			&\theta_t+\frac{2b}{3(a+bt)}\theta
			-\frac{5\Kn b^2\sqrt{\theta}}
			{18\rho d^2\sqrt{\pi}(a+bt)^2}=0.
		\end{aligned}\right.
		\label{eq:hard-sphere-reduced-ns}
	\end{equation}
	For the initial data
	\(
	\rho(t_0)=\rho_0
	\)
	and
	\(
	\theta(t_0)=\theta_0
	\), direct integration of \eqref{eq:hard-sphere-reduced-ns} gives
	\begin{equation}
		\left\{
		\begin{aligned}
			\rho(t)
			&=\frac{\rho_0(a+bt_0)}{a+bt},\\
			\theta(t)
			&=\left\{
			\frac{5\Kn b}
			{12d^2\sqrt{\pi}\rho_0(a+bt_0)}
			+\left[
			\sqrt{\theta_0}
			-\frac{5\Kn b}
			{12d^2\sqrt{\pi}\rho_0(a+bt_0)}
			\right]
			\left(\frac{a+bt_0}{a+bt}\right)^{1/3}
			\right\}^{2}.
		\end{aligned}
		\right.
		\label{eq:hard-sphere-reduced-ns-solution}
	\end{equation}
	Equation \eqref{eq:hard-sphere-reduced-ns-solution} is valid
	on an interval on which $a+bt$ does not vanish or change sign and the displayed
	square-root branch remains positive. It is an exact solution of the
	first-Sonine hard-sphere reduced-NS system \eqref{eq:hard-sphere-reduced-ns}.

	We compare the homogeneous hard-sphere solution with DSMC, starting from a Maxwellian with
	$\rho_0=\theta_0=1$. The spatially homogeneous, axisymmetric solver combines exact axial drift
	with a collision step using a majorant frequency and Strang splitting
	\citep{Cercignani2000,Tcheremissine2005}. Independent replicas provide the ensemble mean and its
	Monte Carlo standard error. This test contains no radial discretization.
	\begin{figure}[H]
		\centering
		\includegraphics[width=0.98\linewidth]{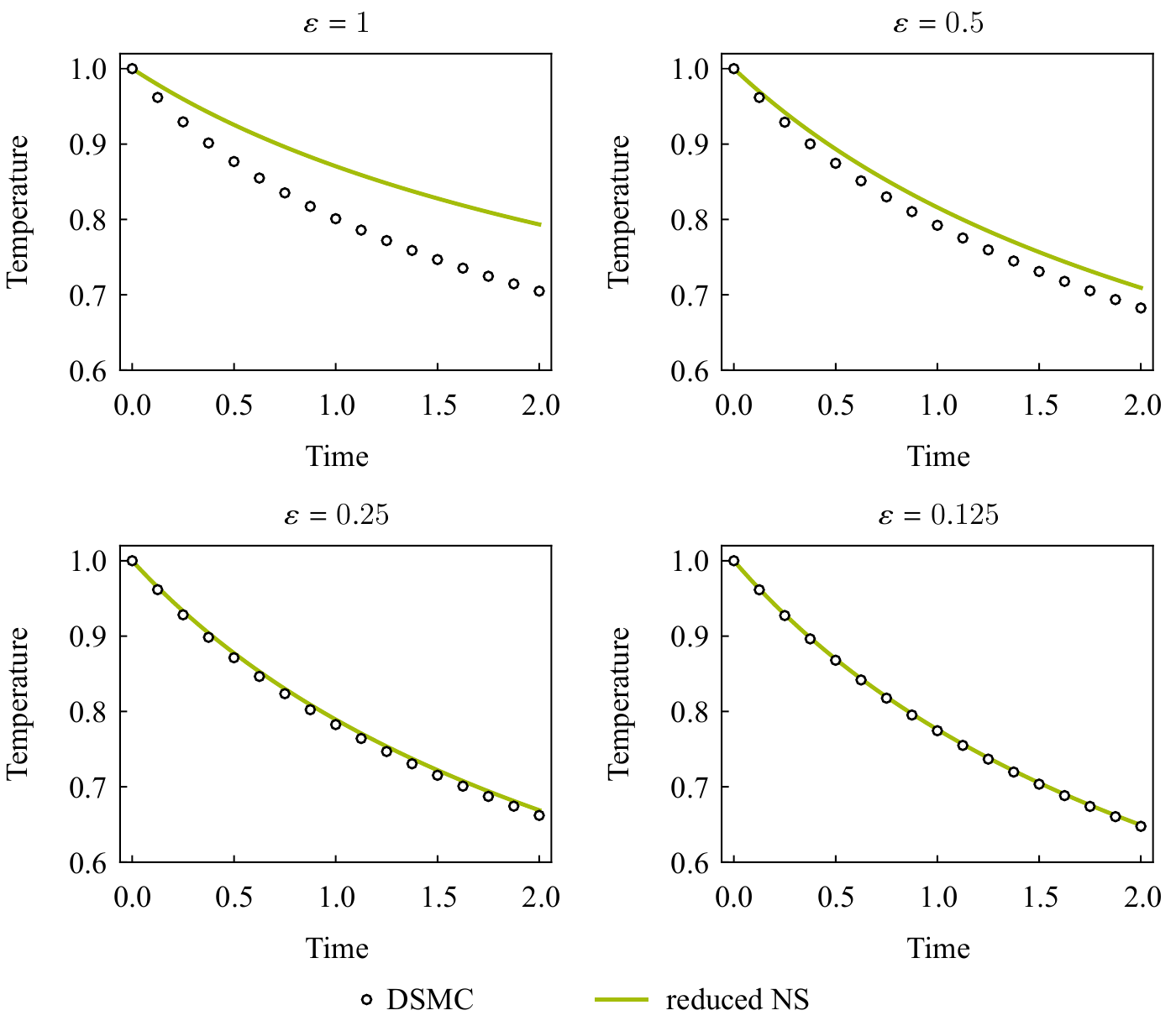}
		\caption{Temperature evolution in homogeneous hard-sphere flow for the
			set $\Kn=1,0.5,0.25,0.125$: first-Sonine reduced NS
			curves and DSMC points.}
		\label{fig:kinetic-temperature}
	\end{figure}
	For a replica with $N$ particles and sampled velocities {$\boldsymbol{W}_{ij}^{(\ell)}$, where $\boldsymbol{W}_{ij}^{(\ell)}$ denotes the reduced velocity $\w$ of particle $i$ in replica $\ell$} at time $t_j$, the
	kinetic temperature is computed after subtracting the instantaneous sample mean:
	\begin{equation}
		\theta_{\mathrm{kin}}^{(\ell)}(t_j)
		=\frac{1}{3N}\sum_{i=1}^{N}
		\left|\boldsymbol{W}_{ij}^{(\ell)}-\overline{\boldsymbol{W}}_j^{(\ell)}\right|^2,
		\quad
		\overline{\boldsymbol{W}}_j^{(\ell)}=\frac{1}{N}\sum_{i=1}^{N}\boldsymbol{W}_{ij}^{(\ell)}.
		\label{eq:dsmc-kinetic-temperature}
	\end{equation}
	The reported kinetic temperature is the mean over replicas. The axial and transverse
	stress diagnostics are formed from the corresponding diagonal components of the
	peculiar second moment and normalized by $\rho\theta_{\mathrm{kin}}$.
	
	Figure~\ref{fig:kinetic-temperature} compares DSMC and the reduced-NS
	prediction for $\Kn=1,0.5,0.25,0.125$, with $a=1$, $b=1/2$, $d=1/2$,
	$t_0=0$, and $\rho_0=\theta_0=1$. Each calculation uses $200{,}000$ particles
	per replica, eight independent replicas, {and a time step $\Delta t=0.005$} up to
	$t=2$. The full interval is shown, including the initial relaxation layer.
	This comparison illustrates the temperature discrepancy at finite Knudsen number $\Kn$, showing that the results reach excellent agreement below $\Kn=0.25$. {Table~\ref{tab:kinetic-temperature-large-kn} lists the endpoint temperatures at $t=2$; the relative difference between the reduced-NS prediction $\theta_{\text{NS}}$ and the DSMC mean decreases $\bar{\theta}_{\text{kin}}$ from approximately $12.6\%$ at $\Kn=1$ to $0.26\%$ at $\Kn=0.125$. The results of reduced Euler $\theta_{\text{E}}$ are also presented here, with a relatively large discrepancy.}

	\begin{table}[H]
		\centering
		\small
		\caption{Endpoint temperatures for the comparison in
			Figure~\ref{fig:kinetic-temperature}. The values are replica means at
			$t=2$. The reduced-NS column uses the first-Sonine closure.}
		\label{tab:kinetic-temperature-large-kn}
		\begin{tabular}{cccc}
			\toprule
			$\Kn$ & $\overline{\theta}_{\rm kin}(2)$
			& $\theta_{\rm NS}(2)$ & $\theta_{\rm E}(2)$ \\
			\midrule
			$1$     & $0.704870$ & $0.793336$ & $0.629961$ \\
			$0.5$   & $0.682488$ & $0.709296$ & $0.629961$ \\
			$0.25$  & $0.662131$ & $0.669040$ & $0.629961$ \\
			$0.125$ & $0.647654$ & $0.649353$ & $0.629961$ \\
			\bottomrule
		\end{tabular}
	\end{table}
	
	\begin{remark}[Maxwell molecules]
		For Maxwell molecules, the homogeneous co-moving branch admits an
		explicit NSF solution. With the non-positive convention for $L$,
		\begin{equation}
			LQ_{33}=\lambda_{02}Q_{33},
			\quad
			\lambda_{02}
			=-\frac{3\pi}{2}\int_{-1}^{1}g(\mu)(1-\mu^2)\,\mathrm d\mu<0,
			\label{eq:maxwell-eigenvalue}
		\end{equation}
		where $g$ is the angular kernel in the
		${\boldsymbol d}$ parametrization with standard sphere area measure.
		The homogeneous NSF equations become
		\begin{equation}
			\rho_t=-\frac{b\rho}{a+bt},
			\quad
			\theta_t
			=-\frac{2b\theta}{3(a+bt)}
			-\frac{8\Kn b^2\theta}
			{9\rho(a+bt)^2\lambda_{02}}.
			\label{eq:maxwell-reduced-ns}
		\end{equation}
		For $\rho(t_0)=\rho_0>0$ and $\theta(t_0)=\theta_0>0$,
		their solution is
		\begin{equation}
			\begin{aligned}
				\rho(t)&=\rho_0\frac{a+bt_0}{a+bt},\\
				\theta(t)&=\theta_0
				\left(\frac{a+bt_0}{a+bt}\right)^{
					\frac23+\frac{8\Kn b}{9\rho_0(a+bt_0)\lambda_{02}}}.
			\end{aligned}
			\label{eq:maxwell-reduced-ns-solution}
		\end{equation}
		These expressions hold while $a+bt$ remains nonzero with fixed sign.
		The constitutive response is exact for Maxwell molecules at NSF order.
		The resulting solution is exact for this macroscopic system, not for
		the Boltzmann equation.
	\end{remark}
	
	\section{Kinetic-to-fluid numerical comparisons}
	We now examine how the first-order constitutive corrections affect the agreement between the reduced kinetic and fluid descriptions. We first consider stress anisotropy in the homogeneous setting, where axial deformation competes with collisional relaxation, and compare the NSF prediction with DSMC results. We then turn to a radially inhomogeneous flow and compare the Euler and NSF solutions against the same kinetic data at a fixed Knudsen number. 
{Together, these comparisons assess the first-order closure for non-equilibrium stresses and spatially varying macroscopic fields.}
	\subsection{Stress anisotropy and the geometry--collision balance}
	\label{sec:stress-anisotropy}
	
	Temperature measures the trace of the pressure tensor and does not determine its
	anisotropy. A directly observable measure of the tensor mode is
	\begin{equation}
		P_\perp=\frac{P_{11}+P_{22}}2,\quad
		\Delta_P=\frac{P_{33}-P_\perp}{p}
		=\frac{3}{2p}\int_{\mathbb R^3}Q_{33}(\cvar)g\,\dd\w.
		\label{eq:stress-anisotropy-observable}
	\end{equation}
	Both $P_{33}$ and the transverse trace are invariant under an active rotation
	about the helical axis, so this centered-moment diagnostic has the same axial
	versus transverse interpretation in the laboratory frame. The Euler
	closure uses an isotropic local Maxwellian, so $\boldsymbol P=p\I$:
	it retains pressure but has no viscous pressure deviation $\boldsymbol\tau$.
	Consequently $\Delta_P=0$, whereas \eqref{eq:nsf-stress-components} gives
	\begin{equation}
		\Delta_P^{\rm NSF}=\frac{\Kn\mu}{p}
		\left(u_{1,r}+\frac{u_1}{r}-\frac{2b}{a+bt}\right).
		\label{eq:stress-anisotropy-general}
	\end{equation}
	Radial deformation can therefore reinforce or oppose the axial contribution.
	
	On the homogeneous branch this reduces to
	\begin{equation}
		\Delta_P^{\rm NSF}=-\frac{2\Kn\mu}{\rho\theta}\frac{b}{a+bt}.
		\label{eq:stress-anisotropy-homogeneous}
	\end{equation}
	For positive pitch, expansion produces a lower axial pressure than transverse
	pressure, while compression reverses the sign. For first-Sonine hard spheres,
	$\Delta_P=-5\Kn b/[8\rho d^2\sqrt{\pi\theta}(a+bt)]$.
	Since $\rho(a+bt)$ is constant along this homogeneous solution, cooling increases
	the magnitude of this prediction through $1/\sqrt\theta$, even though
	$b/(a+bt)$ decreases during expansion. The instantaneous deformation rate alone
	is therefore insufficient to assess departure from equilibrium.
	
\begin{figure}[H]
		\centering
		\includegraphics[width=\linewidth]{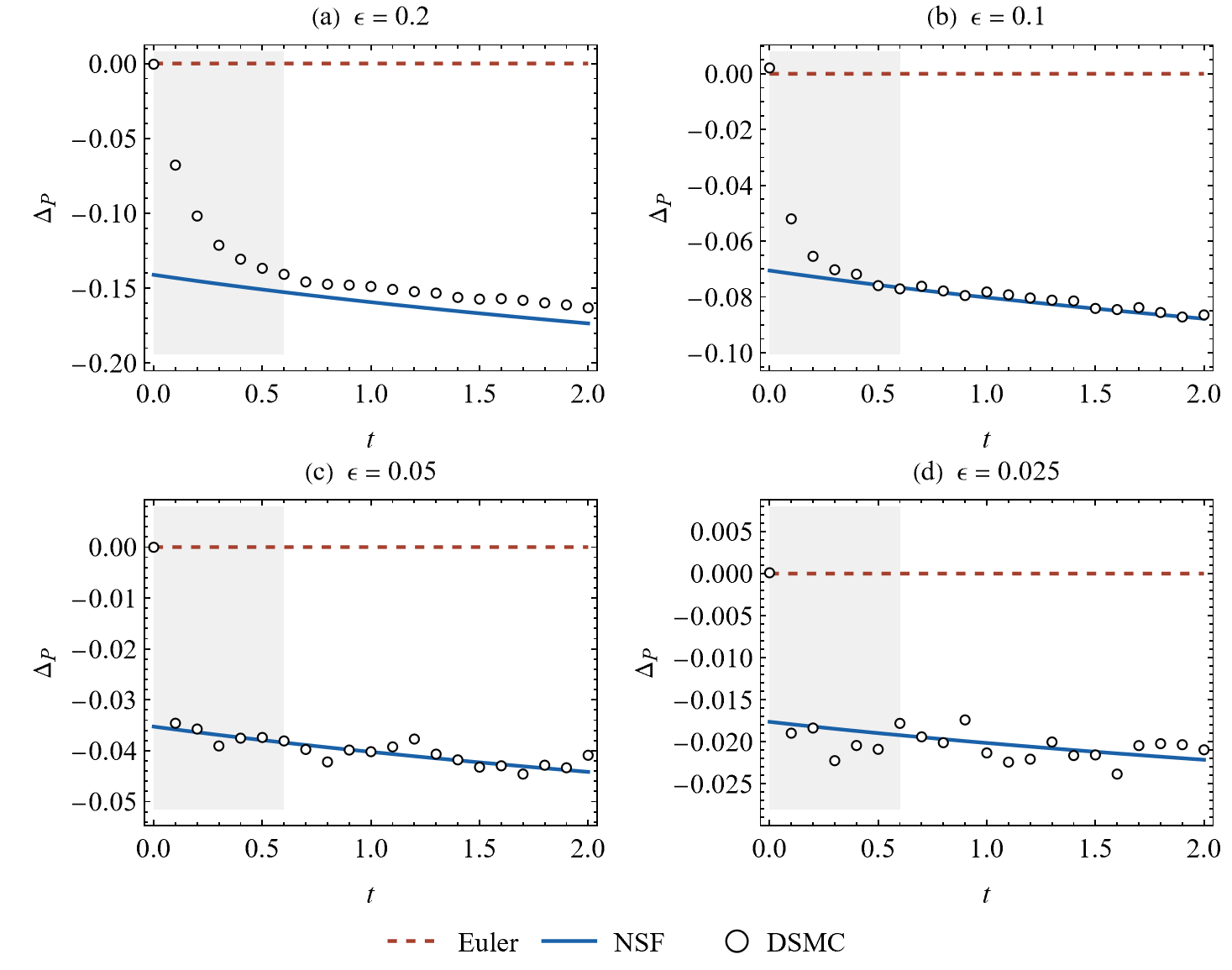}
		\caption{Homogeneous stress anisotropy $\Delta_P$: Euler, NSF and DSMC.
			The shaded interval is excluded from the statistics.}
		\label{fig:stress-anisotropy}
	\end{figure}

	Axial deformation generates the pressure difference, while collisions relax it towards isotropy.
	The NSF law describes the first-order balance between these effects near local equilibrium.
	Figure~\ref{fig:stress-anisotropy} uses the small-$\Kn$ homogeneous
	DSMC ensemble with $a=1$, $b=1/2$, $d=1/2$, $t_0=0$,
	$\rho_0=\theta_0=1$, $t_{\rm end}=2$, and
	$\Kn=0.2,0.1,0.05,0.025$, $200{,}000$ particles per replica,
	eight replicas, and $\Delta t=0.005$.
	The uncertainty of $\Delta_P$ is computed across replicas,
	accounting for the dependence between axial and transverse pressures.
	For a common comparison window $0.6\leq t\leq2$, Table~\ref{tab:stress-common}
	uses all 57 saved times for each Knudsen number.
	\begin{table}[htbp]
		\centering
		\caption{Absolute RMS errors of $\Delta_P$ on the common time window, and
			the RMS of its one-standard-error estimates across eight DSMC replicas.}
		\label{tab:stress-common}
		\begin{tabular}{rrrr}
			\toprule
			$\Kn$ & Euler RMS & NSF RMS & RMS SE\\
			\midrule
			0.2 & 0.153404 & 0.010342 & 0.001178\\
			0.1 & 0.081478 & 0.001424 & 0.001580\\
			0.05 & 0.041311 & 0.001461 & 0.001439\\
			0.025 & 0.020825 & 0.001418 & 0.001407\\
			\bottomrule
		\end{tabular}
	\end{table}
	NSF captures the anisotropy absent from Euler. At $\Kn\leq0.1$, its residual
	is comparable to the sampling-error scale, so these stress data do not resolve
	a second-order remainder. This comparison is descriptive: temporal correlations
	and deterministic discretization errors have not been removed. The initial
	near-Maxwellian DSMC state has nearly zero anisotropy, while the instantaneous
	first-order constitutive stress is nonzero, producing the visible initial
	relaxation layer.
	
	\subsection{A radial NSF comparison at fixed Knudsen number}
	\label{sec:radial-nsf-comparison}

{The general NSF closure is tested against the open-boundary radial kinetic data of Section~\ref{sec:general-euler-validation}, with the same initial fields, $a=1$, $b=1/2$, $d=1/2$, $\Kn=0.025$, and $0.5\leq r\leq2.5$.
The NSF solver uses zero-gradient ghost states at both radial ends.}
		\begin{figure}[H]
		\centering
		\includegraphics[width=0.98\linewidth]{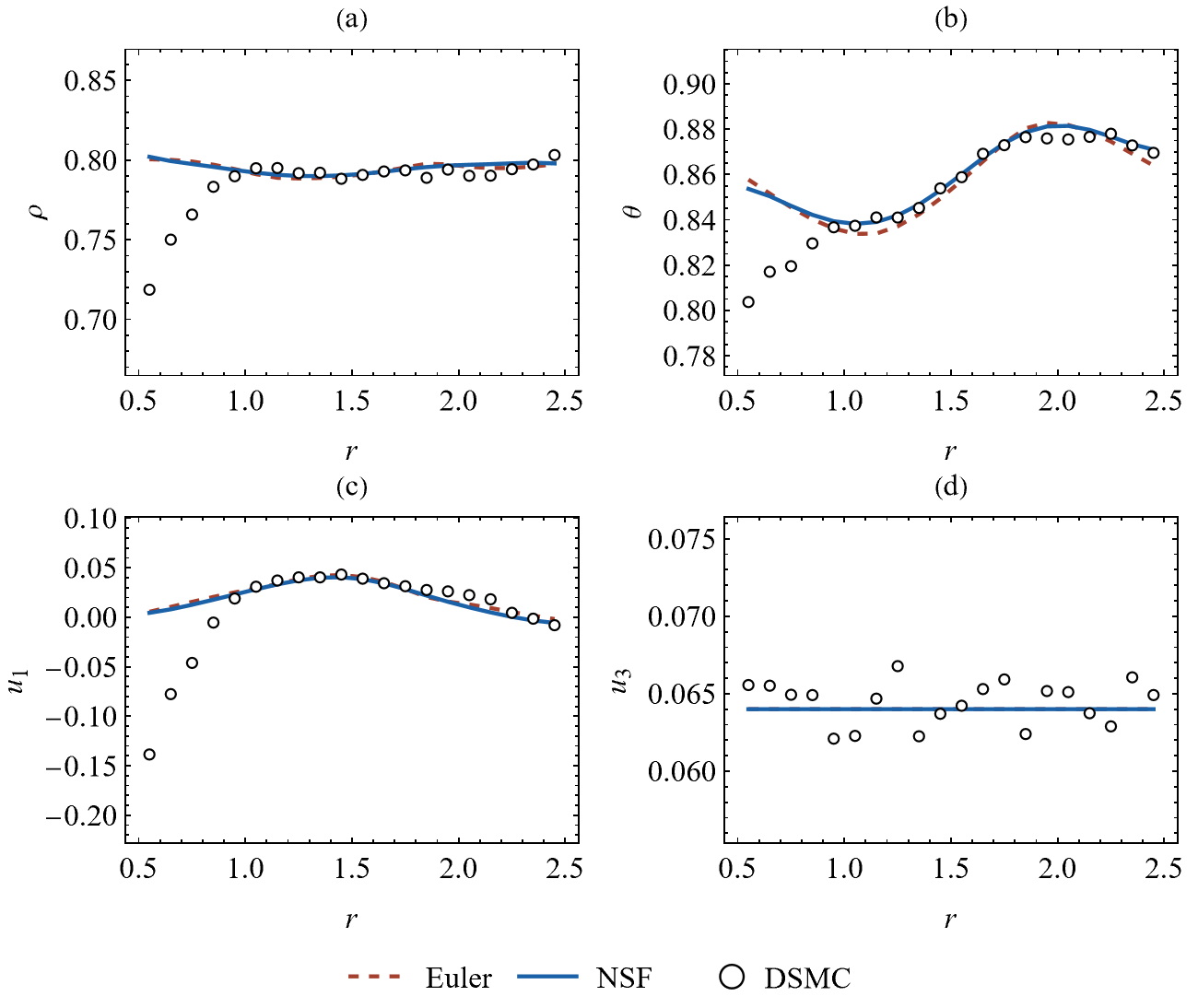}
		\caption{Radial profiles of density, temperature and velocity at $t=0.5$
			and $\Kn=0.025$: (a) $\rho$, (b) $\theta$, (c) $u_1$, and (d) $u_3$. In panel (d), the Euler and NSF curves overlap.}
		\label{fig:radial-nsf-comparison}
	\end{figure}
	
	The NSF discretization extends the area-weighted finite-volume method with centered viscous and
	heat fluxes and the associated geometric stress sources. The lowest Sonine coefficients in
	\eqref{eq:hard-sphere-general-coefficients} are used. Both convective and diffusive stability
	restrictions limit the SSPRK3 time step. 
{Checks recover the Euler operator and the homogeneous analytical solution.
The DSMC mass balance includes the recorded incoming and outgoing particles.}

%
{At $t=0.5$, the temperature comparison on the same 20 conservative observation cells gives the results in Table~\ref{tab:radial-nsf-resolution}.
The quoted time step is an upper bound; the 320-cell NSF calculation takes 2107 steps. The Euler reference uses 640 cells.
At 160 cells, halving the time-step bound changes the final temperature by an absolute RMS of $9.20\times10^{-12}$.
The 160-to-320-cell difference is $4.94\times10^{-4}$, compared with a DSMC RMS standard error of $1.30\times10^{-2}$.
The temperature RMS discrepancies are $1.51\times10^{-2}$ for the 320-cell NSF solution and $1.61\times10^{-2}$ for Euler.
The density relative RMS errors are $2.9239\%$ and $2.9243\%$, respectively, and the radial-velocity absolute RMS errors are $0.04022$ and $0.04090$.
The zero azimuthal mean and spatially constant axial mean of the fluid solutions do not test the corresponding shear responses.
}

{The mean temperature discrepancy is slightly smaller for NSF, but this reduction is small relative to the sampling uncertainty and does not establish a statistically significant improvement.
The different kinetic and fluid boundary treatments
contribute to the near-boundary discrepancies. In other words, to achieve a better agreement at the boundary of a domain, the actual simulated domain needs to be larger.
This interpretation
is supported by a low-resolution domain-extension test, in which
moving the boundaries away reduced the observed DSMC--NSF differences
in the comparison region.}

	\begin{table}[H]
		\centering
		\caption{Final temperature error against the same eight-replica radial DSMC
			ensemble at $\Kn=0.025$. Relative RMS uses the RMS reference temperature.}
		\label{tab:radial-nsf-resolution}
		\begin{tabular}{lrrr}
			\toprule
			Model & Cells & Maximum step & Relative RMS (\%)\\
			\midrule
			Euler & 640 & 0.00025 & 1.87633\\
			NSF & 80 & 0.0005 & 1.74158\\
			NSF & 160 & 0.0005 & 1.75291\\
			NSF & 160 & 0.00025 & 1.75291\\
			NSF & 320 & 0.00025 & 1.75940\\
			\bottomrule
		\end{tabular}
	\end{table}
	
	\section{Discussion and conclusions}
	
	We have developed a reduced kinetic description of a dilute gas under time-dependent helical symmetry, using a continuous invariance ansatz motivated by objective molecular dynamics. The resulting Boltzmann equation retains radial variations and the classical collision operator, while the geometry enters through radial transport, cylindrical curvature and axial deformation. Its formal hydrodynamic approximations yield the reduced Euler and Navier--Stokes--Fourier systems. 
{Comparisons with DSMC solutions connect these fluid descriptions to the
underlying kinetic model, demonstrating the benefits of first-order
corrections in the homogeneous benchmarks and assessing their predictions
for radially varying flow with open boundaries.}
	
	At the Euler level, collisions maintain local equilibrium to leading order, giving an isotropic pressure and adiabatic temperature evolution. The reduced equations show how radial motion, swirl and axial deformation govern the expansion or compression of the gas. In the homogeneous branch, axial extension produces dilution and cooling, whereas compression produces concentration and heating. The radially varying calculations illustrate how spatial differences in expansion generate temperature variations even from an initially uniform temperature. Reconstructing the laboratory velocity further connects the reduced solution to a physical flow field with swirl and axial motion.
	
	The Navier--Stokes--Fourier description captures departures from local equilibrium that are absent from Euler. Axial deformation affects molecular motion differently along and across the axis, while collisions tend to restore an isotropic velocity distribution. This competition produces a pressure difference between the axial and transverse directions. The first-order correction describes this stress anisotropy, the conversion of mean-flow kinetic energy into heat through viscous dissipation, and the redistribution of heat by radial conduction. 
    The homogeneous stress comparisons and the improved radial temperature predictions demonstrate the role of these effects, while the discrepancies at larger Knudsen numbers indicate the limits of the first-order approximation. {Related limitations of the Navier--Stokes description at high deformation rates have also been observed in molecular simulations of Lennard--Jones argon \citep{Pahlani2023HighRate}.}
	
	Beyond these approximation and numerical questions, the reduced equations raise mathematical
	questions concerning long-time behaviour, stability and possible singularity formation. The
	homoenergetic studies of James, Nota and Vel\'azquez
	\citep{JamesNotaVelazquez2019,JamesNotaVelazquez2019Collision,JamesNotaVelazquez2020}
	provide a starting point for investigating asymptotic regimes, while radial dependence introduces
	questions about the stability of homogeneous branches and the interaction of geometric deformation,
	viscous dissipation and heat conduction. The recent finite-time blowup construction reported by
	OpenAI for the three-dimensional incompressible Navier--Stokes equations with smooth forcing
	\citep{OpenAI2026Blowup} provides broader motivation for studying such questions, {particularly given their common helical symmetry}. Its setting
	differs from the present compressible system derived from kinetic theory, so the construction
	does not necessarily imply a corresponding singularity in the reduced equations. The helical reduction
	nevertheless offers a concrete framework for studying asymptotic behaviour and assessing the
	validity of fluid approximations as gradients grow.
	
\section*{Acknowledgment}
We acknowledge financial support from the National Natural Science Foundation of China (Grant Nos. 12588301 and 12472061).
During the preparation of this work the authors used AI-assisted tools (ChatGPT-6 Astra) to help revise and format the manuscript text. After using these tools, the authors reviewed and edited the content as needed and take full responsibility for the content of the publication.

\section*{Declaration of Interests}
The authors report no conflict of interest.

	\bibliographystyle{unsrt} \bibliography{reference_revised}
	
\end{document}